\documentclass[prc,aps,twocolumn,floatfix]{revtex4}

\usepackage[dvips]{graphicx}
\usepackage{epsfig}
\usepackage{pst-plot}
\usepackage{bm}
\usepackage{amsfonts}

\draft
\begin{document}
\title{Generalized contour deformation method in momentum space:
two-body spectral structures and scattering amplitudes.}
\author{G. Hagen}
\affiliation{Department of physics, University of Bergen, N-5000 Bergen, Norway} 
\author{M.~Hjorth-Jensen}
\affiliation{Department of Physics and Centre of Mathematics for Applications, 
University of Oslo, Norway}
\author{J. S. Vaagen}\affiliation{Department of Physics, University of Bergen, 
N-5000 Bergen, Norway} 
\date{\today}

\begin{abstract}
A generalized contour deformation method (GCDM) which combines complex rotation
 and translation in momentum space, is discussed. GCDM gives accurate results 
for bound, virtual (antibound), resonant and 
scattering states starting with a realistic nucleon-nucleon interaction. 
It provides a basis for full off-shell $t$-matrix calculations
both for real and complex input energies. GCDM competes favorably with analytic
continuation. Results for both spectral structures and scattering amplitudes
compare perfectly well with exact values for the separable Yamaguchi potential.
Accurate calculation of virtual states in the Malfliet-Tjon and the realistic 
CD-Bonn nucleon-nucleon interactions are presented. 

GCDM is also a promising method for the computation of in-medium properties such as
the resummation of particle-particle and particle-hole diagrams
in infinite nuclear matter. Implications for in-medium scattering are discussed. 
\end{abstract}

\maketitle

\section{Introduction}
\label{sec:introduction}
The study of two-body resonant structures has a long history in 
theoretical physics, and there exists a variety of methods, described in
textbooks such as \cite{newton,kukulin}. Among the more popular methods 
we have the complex scaling method (CSM) and the method based on 
analytic continuation in the coupling constant (ACCC).

In this work we consider a new approach formulated for integral equations 
in momentum space.  The method is based on 
deforming the contour integrals in momentum space, known as  
the contour deformation or distortion 
method (CDM). It has been shown in Ref.~\cite{afnan1} that a \emph{contour rotation} in 
momentum space is equivalent to a rotation of the corresponding differential equation in
coordinate space. The coordinate space analog is often referred to as the 
\emph{dilation group transformation}, or \emph{complex scaling}. The
\emph{dilation group transformation} was first discussed and formulated in 
Refs.~\cite{abc,abc1}, and was developed to examine the spectrum of the Green's 
function on the second energy sheet. 

Complex scaling in coordinate space has for a long time 
been used extensively in atomic and molecular physics, see
Ref.~\cite{moise}. 
During the last decade it has also been applied in nuclear physics, 
as interest in loosely bound nuclear halo systems has grown, see for 
example Refs.~\cite{csoto, garrido, imante}. Complex scaling 
in coordinate space is usually based on a variational method \cite{moise}, and an 
optimal variational basis and scaling parameters have to be searched for. One of the disadvantages
of the coordinate space approach is that the boundary conditions have to be 
built into the equations, and convergence may be slow if the basis does not 
mirror the physical outgoing boundary conditions well.  

There are several advantages in considering the contour deformation method 
in momentum space. First, most realistic potentials derived from field theoretical 
considerations are given explicitly in momentum space. Secondly, the boundary conditions 
are automatically built into the integral equations. Moreover, 
the Gamow states \cite{kukulin} in momentum space
are non-oscillating and rapidly decreasing, even for Gamow 
states with large widths far from the real 
energy axis,
as opposed to the complex scaled coordinate space counterpart. These states
are represented by  
strongly oscillating and exponentially decaying functions. 
Finally, numerical procedures are
often easier to implement and check. Convergence is easily obtained by just increasing
the number of integration points in the numerical integration. 

The contour deformation method (CDM) formulated in momentum space is not new
in nuclear physics. It was studied and applied in the 1960`s  and 1970`s,  
see for example Refs.~\cite{brayshaw,nuttal,stelbovics}, especially in the field of 
three-body systems. These references
applied a \emph{contour rotation} method in momentum space. By restricting 
oneself to a rotated contour certain limitations and restrictions however appear in the
equations, determined by the analytical structure of the integral kernels and potentials. 
We will study an alternative approach, by considering an extended deformation of the integration 
contour based on rotation followed by translation in the complex momentum plane. This choice 
of contour can be regarded as a special case of the \emph{Berggren class} of contours 
\cite{berggren}. Berggren \cite{berggren} and later Lind \cite{lind} studied various 
completeness relations derived by analytic continuation of the Newton completeness 
relation \cite{newton} to the complex plane. The Berggren completeness 
includes discrete summation 
over resonant as well as bound states. Our choice of contour differ from the
 \emph{Berggren class} of contours in that the contour approaches infinity along complex
rays in the complex $k$-plane 
as opposed to the various contours studied by Lind \cite{lind} which approach 
infinity along the real $k$-axis. 
 By transforming the momentum space Schr\"odinger 
equation onto the rotated followed by translated contour, 
we will show that we are able to expose and explore more of the physical 
interesting area
on the second energy sheet, i.e., the choice of contour enables us to study both
bound, virtual and resonant states. 
If one restricts the deformation to a rotation of the contour,
as studied in Refs.~\cite{brayshaw,nuttal,stelbovics,nuttal1,tikto}, 
we are not able to 
expose virtual states in the
energy spectrum, since the maximum rotation angle does not allow rotation into the
third quadrant of the complex momentum plane. This limitation is sometimes used as an 
argument  for advocating different approaches, such as the ACCC method, see the recent work
of Aoyama  \cite{aoyama}.
By distorting the contour by  
rotation and translation  we 
are able to introduce a new feature to the complex scaling method, namely 
\emph{accurate calculation of virtual states as well as bound and resonant states}. 
Our method represents also an alternative to the \emph{exterior complex scaling} method. 
The \emph{exterior complex scaling} method
was formulated to avoid intrinsic non-analyticities of the potential, and in this way 
calculation of resonances in \emph{non-dilation} analytic potentials are possible, see
Ref.~\cite{moise} and references therein.  
By the 
rotated and translated contour choice the CDM is a preferable method compared to 
all other known methods, such as the ACCC method.

The contour deformation method has also been applied to the solution of the
full off-shell scattering amplitude ($t$-matrix), see Refs.~\cite{afnan1,nuttal, stelbovics,afnan}. 
By rotating the integration contour, an integral equation was obtained with a 
compact integral kernel. This has numerical advantages as the kernel is no longer
singular. As discussed in Ref.~\cite{nuttal}, a rotation of the contour gives certain
restrictions on the rotation angle and maximum incoming/outgoing momentum in
the scattering amplitude. We will again show that our extended choice of contour in momentum 
space avoids all these limitations and that an accurate calculation of the
scattering amplitude can be obtained. 

Thus, the method we will advocate allows us to give 
an accurate calculation of the full energy spectrum. Moreover, it yields  
a powerful method for calculating the full off-shell complex scattering amplitude ($t$-matrix). 
It is also rather straightforward to extend
this scheme to in-medium scattering in e.g., infinite nuclear matter.

In section \ref{sec:formalism} we outline the contour deformation method in momentum space, and
in section \ref{sec:tmatrix}  the spectral representation of the full off-shell $t$-matrix is 
given along with the deformed integration contour. Section \ref{sec:yama} presents 
as a test case  
a simple separable interaction which admits analytical solutions for both the energy spectrum
and the $t$-matrix. Section \ref{sec:results}  gives numerical calculations of energy spectra and
the $t$-matrix by the contour deformation method. We present numerical results for 
the Yamaguchi \cite{yamaguchi}, Malfliet-Tjon \cite{malfliet} 
and the charge-dependent Bonn (CD-Bonn) \cite{machleidt} interactions.
Calculation of virtual states of the CD-Bonn interaction is not known by us to have
been performed previously.

\section{Theoretical framework}
\label{sec:formalism}
We will in the following use natural units $\hbar = c = 1$. 
The two-body momentum space Schr\" odinger equation in a partial wave decomposition reads  
\begin{equation}
\label{eq:eq1}
{k^{2}\over 2\mu}\psi_{nl}(k) + {2\over\pi}\int_{0}^{\infty} 
dq q^{2}V_{l}(k,q)\psi_{nl}(q) = E_{nl}\psi_{nl}(k).
\end{equation}
For the sake of simplicity we assume here that
the interaction is spherically symmetric and local in coordinate space and 
without tensor components 
and/or spin-orbit coupling.  When solving the correspondning equations for a more realistic
nucleon-nucleon interaction below, these degrees of freedom will be accounted for.

The Fourier-Bessel transform of the potential $V_{l}(r)$ in coordinate space is given by
\begin{equation}
\label{eq:eq2}
V_{l}(k,k') = \int_{0}^{\infty}dr r^{2} j_{l}(kr)j_{l}(k'r)V_{l}(r). 
\end{equation}
The momentum space Schr\"odinger equation in Eq.~(\ref{eq:eq1}) (with real momenta) 
corresponds to a hermitian Hamiltonian. The eigenvalues will in this case 
always be real, corresponding to discrete bound states ($E_{nl} < 0 $) and
 a continuum of scattering 
states ($E_{nl} > 0$). The eigenstates form a complete set, and for 
a given partial wave $l$ the completeness relation, more precisely known 
as \emph{resolution  of unity},
can be written \cite{newton} 
\begin{equation}
\label{eq:unity1}
{\bf 1} = \sum _{n}\vert\psi_{nl}\rangle\langle\psi_{nl}\vert + 
{1\over 2}\int_{-\infty}^{\infty} dk \vert\psi_{l}(k)\rangle\langle\psi_{l}(k)\vert,   
\end{equation}
The infinite space spanned by this basis is given by all square integrable functions on the real
energy axis, known as the $L^{2}$ space, which forms a Hilbert space.  
Resonant and virtual states can never be obtained 
by directly solving Eq.~(\ref{eq:eq1}), as it stands. In a sense, one can say that
the spectrum of a hermitian Hamiltonian does not display all information 
about the physical system. 

In this paper we study and explore the resonant and virtual state spectra by the 
contour deformation method. This is essentially a transformation of 
Eq.~(\ref{eq:eq1}) into the lower-half complex $k$-plane. Such a transformation of 
Eq.~(\ref{eq:eq1}) can be obtained by an analytic continuation of the completeness
relation of Eq.~(\ref{eq:unity1}) to the complex $k$-plane. One can consider the integral in 
Eq.~(\ref{eq:unity1}) as an integral over the contour $\Gamma = S + C $, where 
the contour $C$ is defined on the real $k$-axis from $-\infty$ to $ +\infty$ and 
the contour $S$ is given by an 
infinite semicircle in the upper-half complex $k$-plane closing the contour $\Gamma$. 
In Ref.~\cite{lind} completeness relations for various \emph{inversion symmetric} contours
in the complex $k$-plane were derived and discussed. \emph{Inversion symmetric} contours are 
defined by the following: if $z$ is on $C$, then  $-z$ is also on $C$. 
The derivation given in Ref.~\cite{lind} was based on analytic 
continuation by deforming the contour $C$ defined on the real $k$-axis. 
These completeness relations can 
be regarded as a generalization of the Berggren completeness relation \cite{berggren}. 
We will therefore label the \emph{inversion symmetric} contours discussed in \cite{lind}
as the \emph{extended Berggren class} of contours.  
The completeness relation of Berggren was an extension of the completeness relation  
of Eq.~(\ref{eq:unity1}) through the inclusion of a finite sum over discrete resonant states.     
By redefining the completeness relation on distorted contours in the complex $k$ plane, one
can show by using Cauchy's residue theorem that  
the summation over discrete states will in general include bound, virtual and 
resonant states \cite{lind}. The eigenfunctions will form a \emph{biorthogonal} set, and 
the normalization follows  the generalized $c$-product \cite{moise,lind} 
\begin{equation}
\label{eq:norm}
\langle\langle \psi_{nl}\vert \psi_{n'l} \rangle\rangle \equiv 
\langle \psi_{nl}^{*}\vert \psi_{n'l}\rangle = \delta_{n,n'}.
\end{equation}
The most general completeness relation on an arbitrary \emph{inversion symmetric} contour 
$C = C^{+} + C^{-} $ can then be written as 
\begin{equation}
\label{eq:unity2}
{\bf 1} = \sum _{n\in \bf{C}}\vert\psi_{nl}\rangle\langle\psi_{nl}^{*}\vert + 
\int_{C^{+}} dz \vert\psi_{l}\rangle\langle\psi_{l}^{*}\vert,   
\end{equation} 
where $C^{+} $ is the distortion of the positive real $k$-axis, and $C^{-} $ is the 
distortion of the negative real $k$-axis. The symmetry of the integrand has been 
taken into account, that is 
\[
\int_{C^{-}} dz \vert\psi_{l}\rangle\langle\psi_{l}^{*}\vert = 
\int_{C^{+}} dz \vert\psi_{l}\rangle\langle\psi_{l}^{*}\vert .
\]
The summation is over all discrete states (bound, virtual and 
resonant states) located in the domain $\bf{C}$,  
defined as the area above the contour $C$, 
and the integral is over the non-resonant complex continuum defined on $C^{+}$.  
The space spanned by the basis given in Eq.~(\ref{eq:unity2}) 
includes all square integrable functions defined in the domain $\bf{C}$. 
The complete basis could then be used to expand resonant and virtual states defined in  
the region above the distorted contour. Such a complete basis is more 
flexible than a complete basis defined for only real energies. 
From the general completeness relation (\ref{eq:unity2}) 
one can deduce the corresponding eigenvalue problem, $H\vert\psi\rangle = E\vert\psi\rangle $.
The Hamilitonian will in this case be complex and non-hermitian, 
as Gamow and virtual states are included in the spectrum. 

In close analogy with the above discussion on completeness relations, the momentum space
Schr\"odinger equation, see Eq.~(\ref{eq:eq1}), defined on the positive real $k$-axis 
can be continued to the lower-half complex $k$-plane. 
Eq.~(\ref{eq:eq1}) is a \emph{Fredholm} integral equation of the 
second kind. Continuing Eq.~(\ref{eq:eq1}) to the lower-half $k$-plane, the general 
rule that the moving singularities of the integral kernel must not intercept the integration
contour must be obeyed, see for example Ref.~\cite{kukulin}. 
The choice of distorted contour will for each partial wave be based on the 
\emph{a posteriori} knowledge of poles in the scattering matrix. 
By considering the transformed momentum space version of Eq.~(\ref{eq:eq1}), the
choice of contour will in addition be dictated by the analytic structure of the integral kernel 
 and potential. The contour must be chosen in such a way that singularities in the 
potential are located outside the closed integration contour. 

In the following we study the analytic continuation of Eq.~(\ref{eq:eq1}) onto 
two distorted contours $C_{1}^{+}$ and $C_{2}^{+}$. These contours can be regarded 
as a special case of the 
\emph{extended Berggren class} of contours. 
The contour $C_{1}^{+}$ is obtained 
by a phase transformation (rotation) into the lower-half complex $k$-plane while 
the second contour $C_{2}$ will be based on rotation 
followed by translation in the lower-half complex $k$-plane. These contours differ in an
important aspect from the 
\emph{extended Berggren class} of 
contours, in that they approach infinity along complex rays, and not along the real $k$-axis. 
It has previously been assumed as a requirement for the choice of distorted contours
that they approach infinity along the real $k$-axis, see for example Ref.~\cite{betan}.  

\begin{figure}[hbtp]
\begin{center}
\resizebox{8cm}{5cm}{\epsfig{file=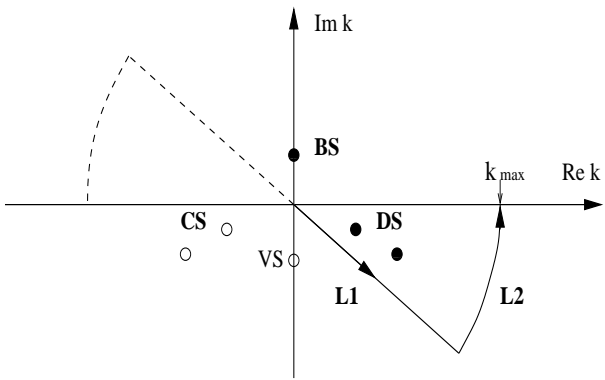}}
\end{center}
\caption{Contour $ C_{1}^{+} = L_{1} + L_{2} $ is given by the solid line, while
the contour $C_{1}^{-} $ is given by the dashed line. The contour $C_{1} = C_{1}^{+}+C_{1}^{-}$ is clearly
\emph{inversion symmetric}. The two body spectrum which is exposed by this contour is marked by 
filled circles $ \bullet $ and the excluded spectrum by open circles $\circ $.
 The full spectrum includes bound states (BS), 
virtual (VS), decay (DS) and capture (CS) resonant states.}
\label{fig:contour1}
\end{figure}

First we consider a contour $C^{+}_{1}$  given by 
two line segments  $L_{1}$ and  $ L_{2}$. Line $L_{1}$ is
 given by $z_{1} = k\exp{(-i \theta)}$ where $k\in [0,k_{max}]$, $L_{2}$ by 
$z_{2} = k_{max} \exp{(-i\theta)}$ ($k$ still real).  
One can easily show that for an exponentially bounded potential in coordinate
space the integral in Eq.~(\ref{eq:eq1}) along the arc $L_{2}$  
will go to zero for $k_{max}\rightarrow \infty $. In this case the contour $C_{1}^{+} $ 
reduces to the line $L_{1}$.  
Fig.~\ref{fig:contour1} shows the contour $C_{1}^{+}$ along with the exposed 
and excluded two-body spectrum in the complex $k$-plane, which this contour choice implies. 
The discrete spectrum consisting of bound, virtual and resonant states corresponds to poles of the 
scattering matrix $S_{l}(k)$ in the complex $k$-plane. 
The contour 
$C_{1}^{+}$ is part of the \emph{inversion symmetric} contour $C_{1} = C_{1}^{+} + 
C_{1}^{-}$, as can be seen in Fig.~\ref{fig:contour1}.  	 
For a potential which is analytic in the region above the contour $C_{1}^{+}$ and below 
the real $k$-axis, we can then derive the transformed
Eq.~(\ref{eq:eq1}) on contour $C_{1}^{+} $
\begin{eqnarray}
\nonumber
\left( \exp{(-2i\theta)}{k^{2}\over 2\mu} - E_{nl}\right) \psi_{nl}(z) + \\ 
\label{eq:eq3}
\exp{(-3i\theta)}{2\over\pi}\int_{0}^{\infty} 
dq q^{2}V_{l}(z,z')\psi_{nl}(z')  = 0,
\end{eqnarray} 
where $z=k\exp{(-i\theta)}$ and $z' = q\exp{(-i\theta)}$. Eq.~(\ref{eq:eq3}) is the
momentum space version of the complex scaled Schr\"odinger equation in
coordinate space, discussed in e.g., Ref.~\cite{afnan1}. 
A rotation in momentum space,
$k\exp{(-i\theta)} $, is equivalent to the complex scaling $ r\exp{(i\theta)} $ 
in coordinate space. The phase-transformation $k\: \rightarrow\: k\exp{(-i\theta)} $
is formally a similarity transformation, see for example Ref.~\cite{brown}. 
The  restriction on the rotation angle $\theta$ for the phase-transformation
$k\: \rightarrow\: k\exp{(-i\theta)} $ is given by the region of
analyticity of the potential. 
It has been shown in Refs.~\cite{abc,abc1} that such a transformation does not alter 
the location of bound states in the 
system; the bound state spectrum is invariant under such transformations. This is also 
clear from the discussion above on completeness relations.  
The continuum is shifted into the lower-half $k$-plane, while resonant states 
will occur as long as they are located above the integration contour.
The transformed Eq.~(\ref{eq:eq3}) represents a non-hermitian Hamiltonian. The eigenvalues
are therefore in general complex, and the corresponding eigenfunctions form 
a \emph{biorthogonal} set with the normalization condition given by Eq.~(\ref{eq:norm}).
   
\begin{figure}[hbtp]
\begin{center}
\resizebox{8cm}{5cm}{\epsfig{file=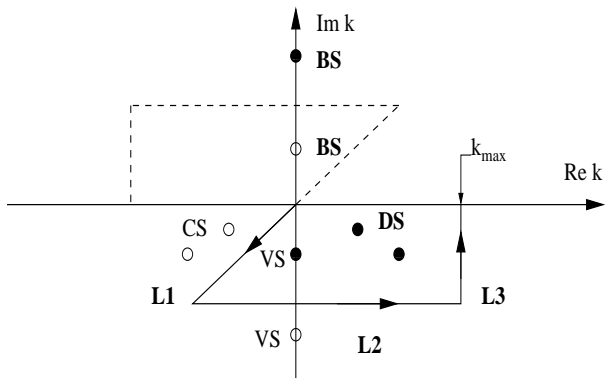}}
\end{center}
\caption{Contour $ C_{2}^{+} = L_{1} + L_{2} + L_{3} $ is given by the solid line, while
the contour $C_{1}^{-} $ is given by the dashed line. The contour $C_{1} = C_{1}^{+}+C_{1}^{-}$ is clearly
\emph{inversion symmetric}. The two body spectrum which is exposed by this contour is marked by
filled circles 
$ \bullet $ and the excluded spectrum by open circles $\circ $. 
The full spectrum includes bound states (BS), 
virtual (VS), decay (DS) and capture (CS) resonant states.  }
\label{fig:contour2}
\end{figure}

Next we consider the contour obtained by rotation followed by translation
in the lower-half complex $k$-plane. The contour $C^{+}_{2}$ consists of three line segments.  
 The line segment $L_{1}$ is given by a rotation $z_{1} = 
k_{1}\exp{(-i\theta)}$ where
 $k_{1} \in [0, b]$, $L_{2} $ is given by a translation 
$z_{2} = k_{2} -ib\sin (\theta)$  where $k_{2} \in [b\cos (\theta), k_{max}]$ and 
 $b$ determines the translation into the lower-half $k$-plane and 
$L_{3}$ by $z_{3} = k_{max} -ic$ where $c\in [b\sin (\theta) ,0]$. 
For $k_{max} \: \rightarrow \: \infty $ the contribution to the integral in Eq.~(\ref{eq:eq1})
along the line segment $L_{3}$  will vanish, and the contour $C_{2}^{+}$ reduces to  
the line segments $L_{1}$ and $L_{2}$.
Fig.~\ref{fig:contour2} shows the contour $C_{2}^{+} = L_{1} + L_{2} + L_{3}$ 
along with the exposed 
and excluded two-body spectrum which this contour choice implies. The contour 
$C_{2}^{+}$ is part of the \emph{inversion symmetric} contour $C_{2} = C_{2}^{+} + 
C_{2}^{-}$ clearly seen in the figure.  	 
We can now transform the momentum space Schr\"odinger equation (\ref{eq:eq1}) onto the distorted contour
$C_{2}^{+} $ given by the lines
$L_{1}$ and $L_{2}$, i.e., $k,k' \: \rightarrow \: z,z' $. 
The integral in Eq.~(\ref{eq:eq1}) will in this case couple the 
complex momenta $z_{1}$ and $z_{2}$. The coupled transformed Schr\"odinger equation then reads
\begin{widetext}
\begin{equation}
\label{eq:eq4}
{1\over2\mu}
\left( \begin{array}{cc}
z_{1}^{2}  & 0 \\
0 & z_{2}^{2} 
\end{array} \right)
\left( \begin{array}{c} 
\psi_{nl}(z_{1}) \\
\psi_{nl}(z_{2})
\end{array}\right)
 + 
{2\over\pi} 
\left( \begin{array}{cc}
\int_{L_{1}} d{z'}_{1}\:{z'}_{1}^{2}V_{l}(z_{1},{z'}_{1}) & 
\int_{L_{2}} d{z'}_{2}\:{z'}_{2}^{2}V_{l}(z_{1},{z'}_{2})  \\
\int_{L_{1}} d{z'}_{1}\:{z'}_{1}^{2}V_{l}(z_{2},{z'}_{1}) & 
\int_{L_{2}} d{z'}_{2}\:{z'}_{2}^{2}V_{l}(z_{2},{z'}_{2})  
\end{array} \right)
\left( \begin{array}{c} 
\psi_{nl}({z'}_{1}) \\
\psi_{nl}({z'}_{2})
\end{array}\right)
= 
E_{nl} 
\left( \begin{array}{c} 
\psi_{nl}(z_{1}) \\
\psi_{nl}(z_{2})
\end{array}\right).
\end{equation}
\end{widetext}
This equation is again  a non-hermitian Hamiltonian. The basis of eigenstates 
forms a \emph{biorthogonal} set, and the normalization is again given by 
Eq.~(\ref{eq:norm}). The completeness relation given in Eq.~(\ref{eq:unity2}), will 
include a discrete sum over bound, virtual and resonant states, and the integration along
$L = L_{1}+L_{2}$ is over the complex non-resonant energy continuum.
  
\begin{figure}[hbtp]
\begin{center}
\resizebox{8cm}{4cm}{\epsfig{file=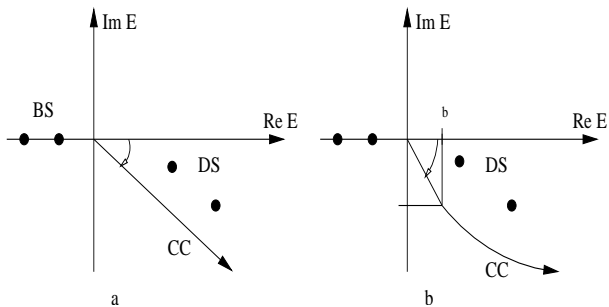}}
\end{center}
\caption{General energy spectra in the complex energy plane for the
 complex transformed
 Schr\"odinger Eqs.~(\ref{eq:eq3}) and (\ref{eq:eq4}), Fig.a and Fig.b,
respectively. 
bound states are denoted by (BS), decay resonant states by (DS) and the complex
continuum by (CC). }
\label{fig:fig2}
\end{figure}

Fig.~\ref{fig:fig2} shows a plot of the energy spectra of the phase-transformed 
Eq.~(\ref{eq:eq3}) and the rotated + translated Eq.~(\ref{eq:eq4}) in the complex 
energy plane, for  a two-body potential supporting bound and resonant states.  

\begin{figure}[hbtp]
\begin{center}
\resizebox{8cm}{4cm}{\epsfig{file=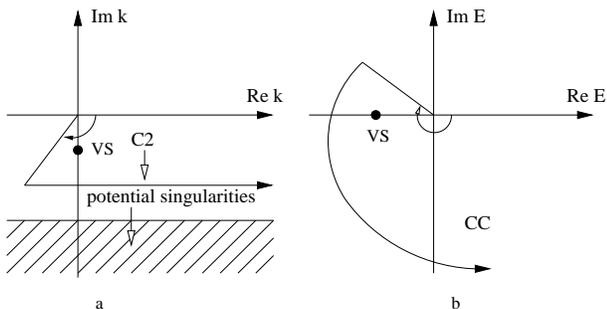}}
\end{center}
\caption{Plot of integration contour $C_{2}^{+}$ where the potential singularities
are located in the marked area in Fig.~a. The rotation angle $\theta$ is greater 
than $\pi /2$ and the integration contour encloses a virtual state (A) in the $k$-plane. 
The corresponding energy spectrum in the complex energy plane in Fig.~b illustrates how the
virtual state (VS) is exposed in the spectrum and the complex non-resonant continuum (CC).} 
\label{fig:fig3}
\end{figure}

Whether one chooses to solve the Schr\"odinger equation on the contour $C_{1}^{+}$ 
or on the contour $C_{2}^{+}$ depends on the problem under consideration. For
potentials which are analytic in the entire lower-half $k$-plane, solving 
along contour $C_{1}^{+}$ is numerically the most straightforward method. 
 In most cases the potential has singularities in 
the lower-half $k$-plane, and solving on the contour $C_{2}^{+}$ enables us to 
avoid the singularities of the potential, while still being able to study 
resonant structures
in the system. Another advantage is that there is no restriction on the 
rotation angle $\theta $ as long as the contour is chosen so that the poles of 
the potential are 
located outside the contour. If a potential is of such an analytic structure 
that we are allowed to rotate into the third quadrant of the complex 
$k$-plane (see Fig.~\ref{fig:fig3}),  virtual states will 
appear in the calculated spectra, as long as the potential supports virtual
states. By solving the Schr\"odinger equation on the distorted contour $C_{2}^{+}$ 
rotated into the third quadrant of the complex $k$-plane, we expose  
a part of the negative imaginary $k$-axis where virtual states may be located, 
while at the same time excluding
a part of the positive imaginary $k$-axis where bound states may be located. 
This reminds us that the 
contour should be chosen relative to the partial wave component under study. This means that a
separate  analysis has to be made for each partial wave. 
  
\section{Two-body $t$-matrix}
\label{sec:tmatrix}
In this section we will discuss different procedures for solving the full
 off-shell $t$-matrix, and hence the full two-body scattering problem. 
We will consider the general mathematical case where the incoming energy is 
allowed to take non-physical values, i.e., the input energy is complex. This has
relevance for nuclear medium studies where the input energy is in general 
complex. Two methods for solving the full off-shell 
$t$-matrix for arbitrary complex energy $\omega $ are outlined. 

The $t$-matrix is defined in operator form by 
\begin{equation}
\label{eq:t1}
t(\omega)  = V + VG(\omega)V,
\end{equation}
or 
\begin{equation}
\label{eq:t2}
t(\omega) =  V + VG_{0}(\omega)t(\omega).
\end{equation} 
Here $\omega$ is the incoming energy, $ G(\omega) $ is the resolvent, 
commonly known as the Green's operator, and $G_{0}(\omega) $ the 
corresponding free Green's operator. In operator form they are defined by
\begin{eqnarray}
G_{0}(\omega) & = & {1\over \omega - H_{0}}, \\ 
\label{eq:greensfunc}
G(\omega) & = & {1\over \omega - H }. 
\end{eqnarray} 
The term $H_{0}$ is the kinetic energy operator and $H$ the full two-body
 Hamiltonian. By expanding the unit operator on a complete set of 
physical eigenstates of $H$ given in Eq.~(\ref{eq:unity1}),
 we can write the Green's operator as
\begin{equation} 
 \label{eq:spectral1}
G(\omega) = \sum _{b} {\vert\psi _{b}\rangle\langle \psi _{b}\vert\over \omega - E_{b}} + \int _{0}^{\infty }dE_{c}{\vert\psi_{c}\rangle\langle\psi_{c}\vert\over \omega - E_{c}}.
\end{equation}	  
This is the spectral decomposition of the Greens's function. 
Here $b$ denotes the discrete bound state spectrum and $c$ the positive
 energy continuum. By projecting $t(\omega)$ on momentum states, and decomposing
 into partial waves, the  $t$-matrix elements $t_{l}(k,k';\omega) $ can be 
expressed as 1-dimensional integral equations. Depending on whether we start
from (\ref{eq:t1}) or (\ref{eq:t2}) we get the \emph{spectral} 
or the  \emph{Fredholm} representation of the $t$-matrix.  The contour 
deformation method will be applied to the \emph{spectral} representation of 
the $t$-matrix while the solution of the \emph{Fredholm} representation of the
 $t$-matrix will
be based on the standard principal value prescription. 

By using Eq.~(\ref{eq:t1}) and Eq.~(\ref{eq:spectral1}) we get the \emph{spectral} representation of the 
$t$-matrix by inserting the expansion of the Green's function on a 
complete set of states (\ref{eq:spectral1}), giving  
\begin{widetext}
\begin{equation}
\nonumber
t_{l}(k,k';\omega) = V_{l}(k,k')  
\label{eq:tmat1}
 + {4\over \pi^{2}}\sum_{\alpha}\int _{0}^{\infty } dq\:\int_{0}^{\infty }dq'\: q^{2}{q'}^{2}
V _{l}(k,q){\psi_{\alpha}(q)\psi_{\alpha}^{*}(q')
\over \omega - E_{\alpha}}V_{l}(q',k'). 
\end{equation}
\end{widetext}
The sum over $\alpha$ implies a discrete sum over bound states and an integration 
over the positive energy continuum. Eq.~(\ref{eq:tmat1}) is analytic along the  
real energy axis, except for poles located at bound state energies and a branch cut along 
the positive energy axis.
In physical two-body scattering the incoming 
energy is defined on the positive real energy axis. In this case 
the integrand in Eq.~(\ref{eq:tmat1}) is singular, and a numerical solution of 
Eq.~(\ref{eq:tmat1}) is highly non-trivial. However, for negative real input energies it
can be solved by standard numerical procedures.    

We are however interested in the $t$-matrix for arbitrary complex energies, since an obvious 
extension of this work is to consider in-medium scattering of two nucleons or
to study the resummation of large classes of many-body diagrams.
In a nuclear medium calculation the self-consistently
determined quasiparticle energies are in general complex.

We can achieve this by an analytic continuation of the contour integrals into the complex
$k$-plane. This represents a deformation of the
integration contour in the complex $k$-plane. See Refs.~\cite{kukulin,nuttal1} 
for validity and mathematical proofs concerning analytic continuation of 
integral equations.

We will consider the solution of
Eq.~(\ref{eq:tmat1}) by integrating over the two contours, $C_{1}^{+}$ or $C_{2}^{+}$,
 discussed in Sec.~\ref{sec:formalism}, resulting in  the following 
expression for $t_{l}(k,k';\omega) $    
\begin{widetext}
\begin{equation}
\nonumber
t_{l}(k,k',\omega) = V_{l}(k,k')  
\label{eq:tmat2}
 + {4\over \pi^{2}}\sum_{\alpha}\int _{C}\int _{C'} dz\:d{z'}
\: z^{2}{z'}^{2}
V _{l}(k,z){\psi_{\alpha}(z)\psi_{\alpha}(z')
\over \omega - E_{\alpha}}V_{l}(z',k'), 
\end{equation}
\end{widetext} 
Here $C$  is either $C_{1}^{+}$ or $C_{2}^{+}$. This applies to $C'$ as well. 
We have expanded the Green's operator (\ref{eq:greensfunc}) on the basis given 
in Eq.~(\ref{eq:unity2}), this
implies that $\alpha $ represents discrete bound, virtual and resonant states and 
a non-resonant complex energy continuum. 
From the mathematical structure of Eq.~(\ref{eq:tmat2}) it is easily seen that
the $t$-matrix is analytic in $\omega $  in the entire complex energy-plane 
except for poles 
located at bound, virtual and resonant state energies. In addition there could
also be singularities in the potential itself. The numerical method for solving 
Eq.~(\ref{eq:tmat2}) is based on matrix diagonalization. We first have to
diagonalize the corresponding Schr\"odinger equation, and then expand the
Green's function on the discretized basis obtained. 

Applying this method enables us to 
obtain $t_{l}(k,k';\omega) $ for both real and complex energies $\omega $. 
The integral becomes non-singular on the deformed contour for real and positive 
input energies $\omega $, resulting in numerically stable solutions for physical
two-body scattering. Eq.~(\ref{eq:tmat2}) can also be considered as an
analytic continuation of Eq.~(\ref{eq:tmat1}) for complex input energies $\omega$,
and stable numerical solution can be obtained for complex energies above the 
distorted contour.  
The limitation of this method is due to that most  potentials
in momentum space have singularities in the complex plane when one argument 
is real and the other is complex.
 By applying contour $C_{1}^{+}$, which is based
on rotation into the complex plane, in most cases there will be restrictions
on both rotation angle ($\theta$) and maximum  incoming and outgoing momentum 
($k,k'$), see for example Ref.~\cite{nuttal}. 

Using contour $C_{2}^{+} $ we can avoid these limitations by 
choosing the integration contour in such a way that the potential 
singularities always will lie outside the integration contour, and therefore
do not give any restriction on rotation angle and maximum incoming and outgoing
momentum.
  
The partial wave decomposition of Eq.~(\ref{eq:t2}) gives the \emph{Fredholm} 
representation, commonly known as the Lippmann-Schwinger equation  
\begin{equation}
\label{eq:tmat3}
t_{l}(k,k';\omega) = V_{l}(k,k')+
{2\over \pi}\int _{0}^{\infty } {dq q^{2}V_{l}(k,q)t_{l}(q,k')\over \omega - E(q) }.
\end{equation}
In physical two-body scattering the input energy is real and positive. In this
case the \emph{Fredholm} integral Eq.~(\ref{eq:tmat3}) has a singular
kernel of Cauchy type. Solving singular integrals can be done 
by Cauchy's Residue theorem, where we 
integrate over a closed contour enclosing the poles. There are two ways of
doing this, either by letting $ z $ lie an infinitesimal distance above the 
real axis, i.e., $ \omega \rightarrow z + i\epsilon $,  or by letting $ \omega $
lie on the real axis. In both cases we must choose a suitable contour enclosing the
 singularity. If we choose the latter position of the singularity,  we get  
 a  Cauchy \emph{principal-value} integral where we integrate up to - but not
 through - the singularity, and a second contour integral,  where the 
contour can be chosen as a semicircle around the singularity. 
Eq.~(\ref{eq:tmat3}) can thus be given in terms of a principal value part and a 
second term coming from integration over the semicircle around the pole. The result is
\begin{widetext}
\begin{equation}
\label{eq:pv1}
 t_{l}(k,k';\omega)  = V_{l}(k,k') + {2\over \pi}
\mathcal{P}\int _{0}^{\infty } {dq q^{2}V _{l}(k,q)t_{l}(q,k')\over \omega - E (q) } 
 \:- \: 2i\mu k_{0}V_{l}(k,k_{0})t_{l}(k_{0},k';\omega).
\end{equation}
\end{widetext} 
By rewriting the principal value integral using the relation
\begin{equation}
\mathcal{P}\int_{0}^{\infty}dk\: {f(k)\over k_{0}^{2} - k^{2}} =  
\int_{0}^{\infty}dk\: {[f(k)-f(k_{0})]\over k_{0}^{2} - k^{2}}, 
\end{equation}
we obtain an equation suitable for numerical evaluation.
Eq.~(\ref{eq:pv1}) can be converted into a set of linear 
equations by approximating the integral by a sum over $N$ 
Gaussian quadrature points $( k_{j}; j = 1,...N  ) $, each weighted by $ w_{j} $ 
The full off-shell $t$-matrix is then obtained by matrix inversion. This
method for solving integral equations is known as the Nystrom method. 
It is numerically effective and stable, except for the rare case when
 the incoming energy $\omega$ coincides with or is very close to one of the 
integration points. 

So far we have only considered physical input energies in 
Eq.~(\ref{eq:tmat3}), but it has been shown in Ref.~\cite{kukulin} 
that the analytically 
continued Lippmann-Schwinger equation to complex energies takes the same form
as Eq.~(\ref{eq:pv1}). By solving the full off-shell $t$-matrix for arbitrary 
complex input energy, we do not have to alter the set of linear equations
obtained for physical energy, the only modification is that the energy is
complex. It should be noted that the Lippmann-Schwinger equation (\ref{eq:tmat2})
can be solved by the contour deformation method as well, giving a compact integral
kernel for positive incoming energies. By contour distortion the principal value 
prescription is avoided, and a numerical solution of Eq.~(\ref{eq:tmat2}) is stable
even for incoming energies coinciding with the integration points.        

\section{An analytically solvable two-body potential}
\label{sec:yama}
We now consider a separable potential given by Yamaguchi \cite{yamaguchi}.
It models $s$- and
$p$-waves.
A separable interaction in momentum space 
is analytically solvable, see for example Ref.~\cite{newton} for a demonstration. 
The Yamaguchi interaction therefore admits analytical solution of the 
full off-shell 
$t$-matrix and the $t$-matrix poles, corresponding to the 
energy spectrum. The Yamaguchi $s$-wave potential supports bound and 
virtual states,
while the Yamaguchi $p$-wave potential supports bound, 
virtual and resonant states. The 
Yamaguchi potential is therefore very useful in 
modelling loosely bound two-body systems
which may have a rich resonant state structure, and for checking the 
numerics in our
calculations of the $t$-matrix and the energy spectrum. 

The $s$-wave Yamaguchi 
potential has the form 
\begin{equation}
V_{0}(k,q) = -\lambda g_{0}(k) g_{0}(q), 
\end{equation}
where
\begin{equation}
g_{0}(k) = {1\over k^{2} + \beta ^{2}}.
\end{equation}
In natural units and mass $2\mu = 1 $ MeV,
 the full off-shell $t$-matrix for the
$s$-wave potential reads \cite{newton}
\begin{equation}
t_{0}(k,q;z) = -\lambda {g_{0}(k)g_{0}(q)\over 1 - \lambda {2\over \pi}
\int_{0}^{\infty}dk\: {k^{2}\over k^{2} - E}{1\over (k^{2}+\beta ^{2})} }.
\end{equation}
The integral in the denominator can be evaluated, giving 
\begin{equation}
\lambda{2\over \pi}\int_{0}^{\infty}dk\: {k^{2}\over  k^{2} - E}
{1\over (k^{2}+\beta ^{2})} = {\lambda\over 2}{(\beta+i \sqrt{E})^{2}\over
\beta (\beta^{2}+E)^{2}}.
\end{equation}
The $s$-wave $t$-matrix in closed form reads
\begin{equation}
t_{0}(k,q;E) = -\lambda {g_{0}(k)g_{0}(q)\over 1 - {\lambda\over 2}  
{(\beta+i \sqrt{E})^{2}\over \beta (\beta^{2}+E)^{2}}}.
\end{equation}
Writing $\kappa = -ik$ where $E = k^{2}$ we can solve for the $t$-matrix poles 
as zeroes of the denominator
\begin{equation}
1 - {\lambda\over 2} {(\beta -\kappa)^{2}\over \beta (\beta^{2}-\kappa^{2})^{2}} = 0.
\end{equation}
Solving for $\kappa$ gives
\begin{equation}
\kappa = -\beta \pm\sqrt{{\lambda\over 2\beta}},
\end{equation}  
and we see that the poles are located along the imaginary $k$-axis. Poles of the $t$-matrix 
located on the positive imaginary $k$-axis represents bound states, while poles located on 
the negative imaginary $k$-axis represents virtual states, giving 
a bound $(\kappa > 0 )$ and a virtual state $( \kappa \leq 0 )$ for $\lambda > 2\beta^{3} $  and for 
$\lambda \leq 2\beta^{3} $ two virtual states. 

\begin{figure}[h]
\begin{center}
\resizebox{6cm}{5cm}{\epsfig{file=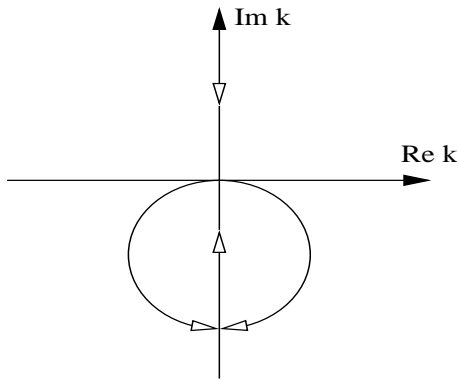}}
\end{center}
\caption{Pole trajectory for the $p$-wave Yamaguchi interaction in the complex
$k$-plane.} 
\label{fig:fig4}
\end{figure}

The separable $p$-wave interaction is given by
\begin{equation}
V_{1}(k,q) = -\lambda g_{1}(k)g_{1}(q),
\end{equation}
where
\begin{equation}
g_{1}(k) = {k\over k^{2} + \beta ^{2} },
\end{equation}
and the $t$-matrix now becomes \cite{newton}
\begin{equation}
t_{1}(k,q;E) = -\lambda {g_{1}(k)g_{1}(q)\over 1 - \lambda {2\over\pi}
\int_{0}^{\infty}dk\: {k^{2}\over k^{2} -E }{k^{2}\over (k^{2}+\beta^{2})^{2}}}.
\end{equation}
Solving again the integral in the denominator gives the $t$-matrix in closed form
\begin{equation}
t_{1}(k,q;E) = -\lambda {g_{1}(k)g_{1}(q)\over 1 - {\lambda\over 2}
{\beta ^{3} + E(3\beta+2i\sqrt{E})\over ( \beta^{2}+ E)^{2}}},
\end{equation}
and solving for the poles gives in terms of $\kappa = -ik$  
\begin{equation} 
\kappa = {\lambda\over 2} - \beta \pm{1\over4}\sqrt{4\lambda^{2}-8\lambda\beta}.
\end{equation}
We see that for $p$-waves the interaction supports bound, virtual and resonant 
states. The bound state condition is 
\begin{equation}
\lambda > {2\beta}. 
\end{equation}
giving in addition a virtual state.  
The interaction has a branchpoint at $k = 0$ where the bound and virtual state
meet and move symmetrically from the imaginary axis into the lower-half $k$-plane
giving capture and decay resonant states. Fig.~\ref{fig:fig4} shows the pole
trajectory for the Yamaguchi $p$-wave interaction. 

\section{Numerical results}
\label{sec:results}
Below we  present numerical calculations for the complete energy spectrum of
the separable Yamaguchi potential \cite{yamaguchi}, for the Malfliet-Tjon
 \cite{malfliet} and the realistic 
charge dependent Bonn (CD-Bonn) \cite{machleidt} nucleon-nucleon 
interactions. We also present calculations of the $t$-matrix for the given
interactions within the formalism outlined above. The analytic structure of the 
various interactions
is of importance for the choice of the integration contour in the complex plane, as
will be discussed.
\subsection{Results for the Yamaguchi potential}
The Yamaguchi potential is analytic in the entire 
complex $k$-plane except for singularities located at $k,k' = \pm i\beta$, arising 
from the separable part of the potential $ 1 /(k^{2}+\beta^{2})$. 
This singularity gives a restriction on the integration contour; 
if we consider the rotated contour $C_{1}^{+}$, the rotation angle $\theta $ has to be less than
$\pi /2$, i.e., $\theta < \pi/2$, in the complex $k$-plane. This contour choice
will not be able to give the virtual states in the calculated energy spectrum.
The contour $C_{2}^{+}$ could on the other hand be chosen to lie above the singularity, i.e., 
$ b\sin(\theta) < \beta $. By this choice there is no restriction on the 
rotation angle $\theta$ and we can rotate into the third quadrant 
of the complex $k$-plane. By this procedure virtual states can be included in 
the calculated energy spectrum. 

\begin{figure}[hbtp]
\begin{center}
\resizebox{6cm}{5cm}{\epsfig{file=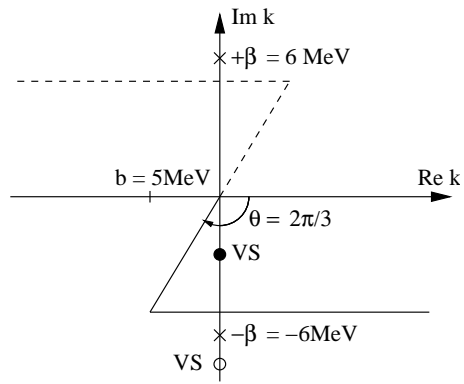}}
\end{center}
\caption{Excluded ($\circ $) and included ($\bullet $) spectrum for the $s$-wave Yamaguchi potential, with 
$\beta = 6$ MeV and $\lambda $ chosen to give only virtual states (VS) in the full spectrum.
The calculation of the spectrum for this contour choice is given in Table~\ref{tab:yama1}. 
Potential singularites are given by ($\times$) in the figure.}
\label{fig:fig6}
\end{figure}

In Table~\ref{tab:yama1} we present numerical versus exact values for the
$s$-wave virtual states in the Yamaguchi potential, using contour $C_{2}^{+}$.
The contour $C_{2}^{+} $ was rotated $\theta = 2\pi /3 $ into the third
quadrant and translated $c = 5\sin(2\pi/3) \approx 4.3301 $ MeV in the lower-half
complex $k$-plane. Fig.~\ref{fig:fig6} shows the contour along with the excluded spectrum and 
the singularities of the $s$-wave Yamaguchi potential giving restrictions on the
contour choice. The parameter $\beta $ was held fixed, and equal to $\beta = 6$ MeV.
The potential depth $\lambda $ was chosen to give only virtual states. The 
contour choice exposes one virtual state while one virtual state will be excluded
from the calculated spectrum. The calculations used  $N1 = 50  $ 
integration points along line $L_{1} $ and $N2 = 50 $
 integration points along $L_{2}$

\begin{figure}[hbtp]
\begin{center}
\resizebox{6cm}{5cm}{\epsfig{file=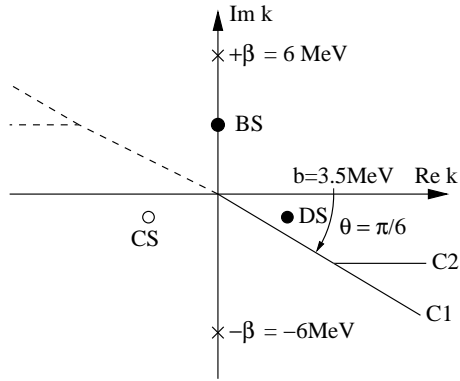}}
\end{center}
\caption{Excluded ($\circ $) and included ($\bullet $) spectrum for the $p$-wave Yamaguchi potential, with 
$\beta = 6$MeV and $\lambda $ chosen to give bound (BS), virtual (VS), capture (CS) and decay (DS) states 
in the full spectrum.
The calculations of the spectrum for contour $C_{1}^{+}$ ($C1$) and contour $C_{2}^{+} $ ($C2$) are 
given in Tables~\ref{tab:tab1} and \ref{tab:tab2}, respectively. 
Potential singularites are given by ($\times$) in the figure.}
\label{fig:yama2}
\end{figure}

In Table ~\ref{tab:tab1} and Table~\ref{tab:tab2} we present numerical versus 
exact values of the $p$-wave energy spectra, calculated using contour 
$C_{2}^{+}$ and $C_{1}^{+}$. The calculations used $N1 = 50  $ 
integration points along line $L_{1} $ and $N2 = 50 $
 integration points along $L_{2}$ for contour $C_{2}^{+}$. For the contour 
$C_{1}^{+}$ a number of $N = 100 $ integration points was used. The rotation angle is the same for
both contours and equal to $ \theta = \pi /6 $. The parameter $\beta $ in the 
potential was held fixed, and equal to $\beta = 6$ MeV. The translation parameter of 
contour $C_{2}^{+}$ is given by $ b = 3.5$ MeV. 
Fig.~\ref{fig:yama2} shows the contour along with the excluded and included spectrum
for the contour $C_{1}^{+}$ and the contour $C_{2}^{+}$, 
along with the singularities of the $p$-wave Yamaguchi potential giving restrictions on the
contour choice.

\begin{table}[hbtp]
\begin{tabular}{rrr}
  &\multicolumn{1}{c}{Exact} & \multicolumn{1}{c}{Contour} $ C_{2}^{+} $ \\
\hline
$\lambda $ & \multicolumn{1}{c}{Real} & \multicolumn{1}{c}{Real} \\
\hline
200 &	-3.67687178 & -3.67687178 \\
250 &	-2.06107759 & -2.06107759 \\
300 &	-1.00000000 & -1.00000000 \\
350 &	-0.35925969 & -0.35925969 \\
390 &	-0.08947449 & -0.08947449 \\
\hline
\end{tabular}
\caption{Numerical calculations for the Yamaguchi $s$-wave 
virtual states versus exact values. 
Energies are given in units of MeV. 
The numerical calculations were obtained using the
contour deformation method (CDM) along contour $C_{2}^{+}$. }
\label{tab:yama1}
\end{table}
\begin{table}[hbtp]
\begin{tabular}{rrrrr}\hline
 & \multicolumn{2}{c}{Exact} & \multicolumn{2}{c}{Contour $ C_{2}^{+} $}  \\
\hline
$\lambda $ & \multicolumn{1}{c}{Real} & \multicolumn{1}{c}{Imag.}&  \multicolumn{1}{c}{Real} &  \multicolumn{1}{c}{Imag.}  \\
\hline
13 &   -5.30277586 &  0 &  	  -5.30277586 & 0. \\ 
12.6 & -2.80486369 &  0 &         -2.80486369 & 0. \\
12.2 & -0.77620500 &  0 &         -0.77620500 & 0. \\
11.8 & 	0.57999998 & -0.15362291 & 0.57999998 & -0.15362291 \\
11.7 & 	0.85500001 & -0.28102490 & 0.85500001 & -0.28102490  \\
11.5 &	1.37500000 & -0.59947896 & 1.37500000 & -0.59947896  \\
11.0 & 	2.50000000 & -1.65831244 & 2.50000000 & -1.65831244  \\
\hline
\end{tabular}
\caption{Numerical calculations for the Yamaguchi $p$-wave 
bound and resonant energy spectra versus exact values. 
Energies are given in units of MeV. 
The numerical calculations were obtained using the
contour deformation method (CDM) along contour $C_{2}^{+}$. }
\label{tab:tab1}
\end{table}

 \begin{table}[hbtp]		
\begin{tabular}{rrrrr}\hline
 & \multicolumn{2}{c}{Exact} & \multicolumn{2}{c}{Contour $ C_{1}^{+} $}  \\
\hline
$\lambda $ & \multicolumn{1}{c}{Real} & \multicolumn{1}{c}{Imag.}&  \multicolumn{1}{c}{Real} &  \multicolumn{1}{c}{Imag.}  \\
\hline
13 &   -5.30277586 &  0 &  	  -5.30277563 & 0. \\ 
12.6 & -2.80486369 &  0 &         -2.80486362 & 0. \\
12.2 & -0.77620500 &  0 &         -0.77620499 & 0. \\
11.8 & 	0.57999998 & -0.15362291 & 0.57999998 & -0.15362291 \\
11.7 & 	0.85500001 & -0.28102490 & 0.85500001 & -0.28102490  \\
11.5 &	1.37500000 & -0.59947896 & 1.37500000 & -0.59947896  \\
11.0 & 	2.50000000 & -1.65831244 & 2.50000000 & -1.65831244  \\
\hline
\end{tabular}
\caption{Numerical calculations for the  Yamaguchi $p$-wave 
bound and resonant state energy spectra versus exact values.
Energies are given in units of MeV. The numerical calculations were
 obtained using the
contour deformation method (CDM) along contour $C_{1}^{+}$. }
\label{tab:tab2}
\end{table}
     
\begin{figure}[hbtp]
\begin{center}
\resizebox{7cm}{7cm}{\epsfig{file=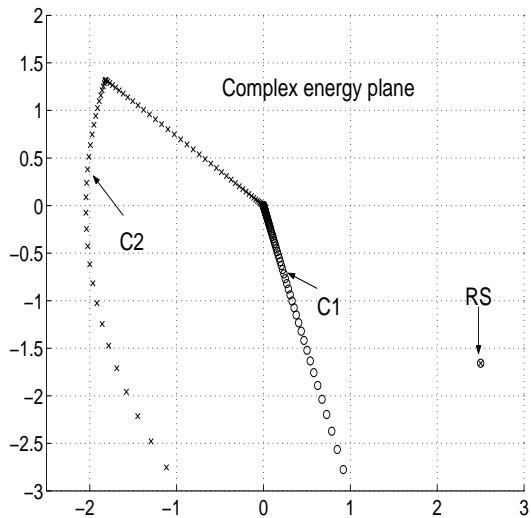}}
\end{center}
\caption{Calculated energy spectra for the $p$-wave Yamaguchi potential with
parameters $\lambda = 11$, $\beta = 6$ using the contours 
$C_{1}^{+}\: (C1),C_{2}^{+}\: (C2)$. The discretized non-resonant continuum lies along the contours, 
and a resonant state (RS) appears in the energy spectra. }
\label{fig:fig7}
\end{figure}

The results for the virtual states for the $p$-wave Yamaguchi potential 
using contour $C_{2}^{+} $,
are given in Table~\ref{tab:tab3}.
\begin{table}[hbtp]
\begin{tabular}{rrr}
  &\multicolumn{1}{c}{Exact} & \multicolumn{1}{c}{Contour} $ C_{2}^{+} $ \\
\hline
$\lambda $ & \multicolumn{1}{c}{Real} & \multicolumn{1}{c}{Real} \\
\hline
13 &    -1.69722438 &	-1.69722438 \\
12.6 &	-1.15513635 &	-1.15513635 \\
12.2 & 	-0.46379500 &	-0.46379500 \\
12.1 &	-0.25000000 &	-0.25000000 \\
\hline
\end{tabular}
\caption{Numerical calculations for the  Yamaguchi $p$-wave 
virtual state energy spectra versus exact values. 
Energies are given in units of MeV.
The numerical calculations were obtained using the
contour deformation method (CDM) along contour $C_{1}^{+}$ with $\theta > \pi /2$.
See Fig.~\ref{fig:fig6} for a plot of the contour $C_{2}^{+}$. }
\label{tab:tab3}
\end{table} \\
We conclude that there is no difference between  the numerically calculated and the 
exact bound, virtual and resonant energy spectra. To see a difference
for the number of integration points used in the calculations we have to include 
up two 12 significant digits. Using double precision we get for the $p$-wave
($\lambda = 13 $) virtual state the exact value $E = -1.697224362268005 $ MeV while
the numerical result in double precision is $ E = -1.697224362276472 $ MeV. 
The calculated results for resonant states with large widths, i.e., far from
 the
real energy axis agree
perfectly with exact values. This illustrates the advantage of the contour 
deformation method in momentum space as opposed to the complex scaling 
analog in coordinate space, where convergence of these states will 
expectedly be slower due to the slowly decreasing oscillating wavefunctions. 
Fig.~\ref{fig:fig7} gives a plot of the calculated energy spectrum for the
$p$-wave Yamaguchi potential,
using contours $C_{1}^{+}$ and $C_{2}^{+}$. Contour $C_{1}^{+}$ is rotated by $\theta = 
2\pi/5 $ in the energy plane while contour $C_{2}^{+} $ is rotated into the third 
quadrant of the complex $k$-plane by $\theta = 3\pi/5 $ giving a rotation of $\theta = 6\pi/5 $
 in the energy plane. The resonant state appears at the same
location for both contours. 

Next we present a calculation of the $t$-matrix elements $t_{l}(k,k;E)$ for 
both real and complex input energy 
(see Table ~\ref{tab:tab4} and Table~\ref{tab:tab5} ) for the $p$-wave 
Yamaguchi potential. 
The $t$-matrix is calculated by applying the
spectral representation on the deformed contour $C_{2}^{+}$. The contour is 
rotated into the third quadrant of the complex $k$-plane and the difference
between numerical calculations and exact values is of the same order as for the
calculations of the energy spectrum. The rotation, translation and number
of integration points were the same as above.
\begin{table}[htbp]
\begin{tabular}{rrrrr}\hline
 & \multicolumn{2}{c}{Exact} & \multicolumn{2}{c}{Contour $ C_{2}^{+} $}  \\
\hline
$k $ & \multicolumn{1}{c}{Real} & \multicolumn{1}{c}{Imag.}&  \multicolumn{1}{c}{Real} &  \multicolumn{1}{c}{Imag.}  \\
\hline
 1.5 & 0.102303714 & -0.061759732 & 0.102303714 & -0.061759732 \\
 3 &   0.295657724 & -0.178485632 & 0.295657724 & -0.178485632 \\
 4 &   0.425747126 & -0.257019311 & 0.425747126 & -0.257019311 \\
 6 &   0.461965203 & -0.278883785 & 0.461965203 & -0.278883785  \\
 7.5 & 0.439705133 & -0.265445620 & 0.439705133 & -0.265445620 \\
 9  &  0.393627137 & -0.237628803 & 0.393627137 & -0.237628803 \\
 10 &  0.342892975 & -0.207001090 & 0.342892975 & -0.207001090 \\
\hline
\end{tabular}
\caption{Calculation of the $p$-wave ($l=1$)
$t_l(k,k,E) $ for real momenta $k$ and input energy 
$E = 5 $ MeV. $t_l(k,k,E) $ is given in units of MeV$^{-2}$.
The interaction parameters are $\lambda = 12.5$, $\beta = 6$ MeV, the rotation
angle is  $\theta = 2\pi/3 $ while the translation is given by $b= 3.5$ MeV.
The number of integration points was $N1 = N2 = 50 $}
\label{tab:tab4}
\end{table}
\begin{table}[htbp]
\begin{tabular}{rrrrr}\hline
 & \multicolumn{2}{c}{Exact} & \multicolumn{2}{c}{Contour $ C_{2}^{+} $}  \\
\hline
$k $ & \multicolumn{1}{c}{Real} & \multicolumn{1}{c}{Imag.}&  \multicolumn{1}{c}{Real} &  \multicolumn{1}{c}{Imag.}  \\
\hline
1.5 & 0.094043918 &  -0.030566057 & 0.094043918 & -0.030566057 \\
3   & 0.271786928  &-0.088335901  &0.271786928  &-0.088335901\\
4.5 & 0.391373158  &-0.127203703  &0.391373158  &-0.127203703\\
6   & 0.424667060  &-0.138024852  &0.424667060 & -0.138024852\\
7.5 & 0.404204220 & -0.131374046  &0.404204220  &-0.131374046\\
9   & 0.361846477 & -0.117606975 & 0.361846477 & -0.117606975\\
10.5& 0.315208495 & -0.102448739 & 0.315208495  &-0.102448739\\
\hline
\end{tabular}
\caption{Calculation of the $p$-wave 
$t_{l}(k,k,E) $ for real momenta $k$ and complex input 
energy  $E = 5 - 2i $ MeV. $t_l(k,k,E) $ is given in units of MeV$^{-2}$.
Interaction parameters are $\lambda = 12.5$, $\beta = 6$ MeV ,
the rotation angle is  $\theta = 2\pi/3 $ while the translation is given 
by $b= 3.5$ MeV. The number of integration points was $N1 = N2 = 50 $.} 
\label{tab:tab5}
\end{table} 
\subsection{Results for the Malfliet-Tjon interaction}
\label{subsec:malfliet}
The Malfliet-Tjon interaction \cite{malfliet}  is a
superposition of Yukawa terms. This interaction resembles the form of a
 realistic nucleon-nucleon interaction with attractive and repulsive parts in the
${}^{1}S_{0}$ channel. It 
can be fitted to reproduce the ${}^{1}S_{0}$ phase shift in nucleon-
nucleon scattering rather well. 
The interaction in coordinate representation is given by 
\begin{equation}
\label{eq:MF1}
V(r) = V_{A}{\exp{(-\mu_{A}r)}\over r} + V_{B}{\exp{(-\mu_{B}r})\over r}.
\end{equation}
This interaction will support bound and virtual states for $s$-waves, while for 
higher angular momentum resonances will appear. It 
is known that the ${}^{1}S_{0}$ channel in the nucleon-nucleon 
interaction supports a virtual state near 
the scattering threshold. 

We will only consider $s$-waves and calculate the
energy spectrum using contour $C_{2}^{+}$ rotated into the third quadrant of the 
complex $k$-plane. 
First we consider the analytic structure of 
the Malfliet-Tjon interaction in momentum space. The Fourier-Bessel transform of
(\ref{eq:MF1}), for the $s$-wave, is
\begin{eqnarray}
\nonumber
V_{0}(k,k') & = & V_{A}{1\over 4 kk'}\ln \left( {(k+k')^{2}+\mu_{A}^{2}\over    
(k-k')^{2}+\mu_{A}^{2}}\right) \\ 
\label{eq:MF2}
& + & V_{B}{1\over 4 kk'}\ln 
\left( {(k+k')^{2}+\mu_{B}^{2}\over (k-k')^{2}+\mu_{B}^{2}}\right).
\end{eqnarray}
By analytic continuation of the interaction (\ref{eq:MF2}) to the complex 
$k$-plane, there will be singularities in the interaction for 
\begin{equation}
\label{eq:sing}
(z - z')^{2}+\mu_{A,B}^{2} = 0,
\end{equation}
and 
\begin{equation}
\label{eq:sing1}
(z + z')^{2}+\mu_{A,B}^{2} = 0.
\end{equation}
Eq.~(\ref{eq:sing}) is only satisfied if one of the arguments is real and the other
complex. For $z' = k' $ real, we have singularities for $\Re[z] = \mp k' \: 
\wedge \: \Im[z] = \pm\mu$.
Eq.~(\ref{eq:sing1}) is satisfied for $ ( z + z' ) = \pm i \mu_{A,B} $, which gives 
$\Re[z] = \Re[z'] \: \wedge \: \Im[z]= \pm\mu_{A,B} -\Im[z'] $.
 Solving the eigenvalue problem by the contour 
deformation method on the purely rotated contour $C_{1}^{+}$, singularities appear along the 
imaginary axis given by $\Im[z] =  \pm\mu_{A,B} -\Im[z'] $.
This gives a restriction on rotation angle, $\theta < \pi/2 $. 
This does not allow a rotation into the third quadrant of the complex
$k$-plane, and calculation of virtual states is not possible by this contour choice.    
This problem is resolved by solving the eigenvalue problem on the rotated + translated
contour $C_{2}^{+}$. For a rotation angle $\theta \geq \pi/2 $ singularities appear on the contour $C_{2}^{+} $ for  
$\Re[z] = \Re[z']\: \wedge \: \Im[z] + \Im [z'] = \pm\mu_{A,B} $. By imposing a lower bound
on the translated line segment $L_{2}$, given by $c < \mu_{A,B} / 2$, we avoid
all singularities. By this choice, rotation into the third quadrant of the complex $k$-plane
is allowed, and an exploration of virtual states is possible. 

If we want to extract the $t$-matrix along the real $k$-axis by the contour
deformation method, see Sec.~\ref{sec:tmatrix}, singularities appear when Eq.~(\ref{eq:sing}) is satisfied. 
If we integrate along contour $C_{1}^{+} $ there will always be a singularity on the 
contour given by 
\begin{equation}
z = k_{max} - i\mu,
\end{equation}
where
\begin{equation}
k_{max} = \mu /\tan(\theta ),
\end{equation}
and $\mu = min[\mu_{A},\mu_{B}] $. 
For $k,k' > k_{max} $ the contour $C_{1}^{+}$ will 
pass through the singularity of the interaction and Cauchy's integral theorem cannot
be applied. However, if the interaction, $V_{l}(k,k') $, is approximately zero for $k,k' > k_{max} $
integrating along contour $C_{1}^{+} $ can be done as long as 
\begin{equation}
\theta < \arctan \left( \mu/ k_{max} \right ).
\end{equation}
In this sense we may call $k_{max} $ the cutoff momentum. 
This choice may cause numerical unstable solutions for small values of momenta, 
since the rotated contour may lie very close to the real $k$-axis where the
integral kernel is singular.  
The same conclusion has already been pointed out in Ref.~\cite{nuttal}.

\begin{figure}[hbtp]
\begin{center}
\resizebox{7cm}{5cm}{\epsfig{file=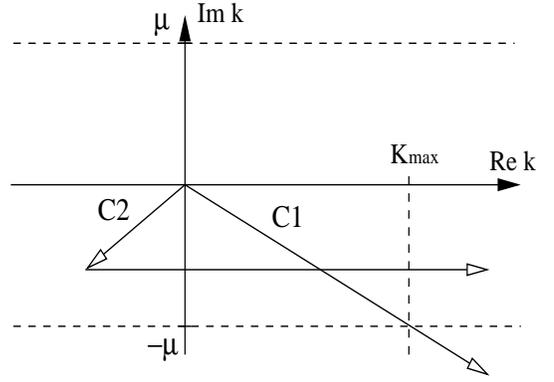}}
\end{center}
\caption{Illustration of potential singularities for the Malfliet-Tjon interaction $V_{l}(k,z')$, 
where $k$ is real  and $z'$ complex. Restriction on the integration contours $C_{1}^{+} \: (C1)$ 
and $C_{2}^{+} \: (C2)$
is clearly seen for given cutoff momentum $k_{max } $ ($V_{l}(k,k') \approx 0 $ for 
$k,k' > k_{max} $). }
\label{fig:fig8}
\end{figure}

Here we see the advantage of integrating along contour $C_{2}^{+}$; Not only 
will it be able to reproduce virtual states in the spectra, but it will also give
accurate calculations of the $t$-matrix for real incoming and outgoing momentum.
It will always  be possible to choose a contour $C_{2}^{+}$ lying above the nearest singularity $z = \Re[k] - 
i\mu $, implying no restriction on rotation angle $\theta $ irrespective of cutoff 
momentum $k_{max} $. Fig.~\ref{fig:fig8} gives an illustration. 

\begin{figure}[hbtp]
\begin{center}
\resizebox{7cm}{6cm}{\epsfig{file=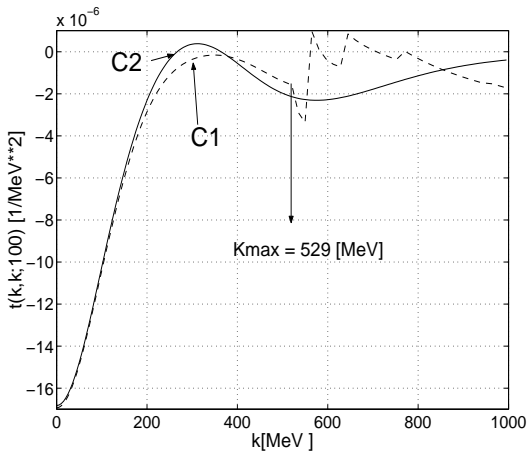}}
\end{center}                                           
\caption{Plot of $t$-matrix elements for the $s$-wave Malfliet-Tjon interaction
using the contour $C_{1}^{+}$ (dashed line) and the contour $C_{2}^{+}$ (solid line).
The potential singularity along contour $C_{1}^{+}$ is clearly displayed, and
located at $k_{max} = 529.77$ MeV.}
\label{fig:fig9}
\end{figure}

A calculation of resonant and virtual states in the Malfliet-Tjon interaction is 
given in Ref.~\cite{elster}. The $t$-matrix poles were calculated by analytic continuation of the
$t$-matrix to the second energy sheet, and thereby searching for poles. In Table~\ref{tab:tab6} 
a calculation of the virtual neutron-proton $s$-wave states with reduced mass
$M_{np} = 938.926$ MeV and 
$ \hbar = c = 1 $ for interaction parameters $ V_{A} = 7.291$ MeV, $\mu_{A} = 613.69$ MeV, 
$\mu_{B} = 305.86$ MeV  held fixed while $V_{B} $ was varied as given in Table~\ref{tab:tab6}. 
The rotation is $\theta = 
2\pi/3 $ and the translation $c = 100\sin(2\pi/3)$ MeV. The convergence is illustrated by 
increasing the number of integration points. 
\begin{table}[hbtp]	
\begin{tabular}{rrrrr}
\hline
\multicolumn{1}{c}{} & \multicolumn{3}{c}{CDM} & \multicolumn{1}{c}{Ref.\cite{elster}} \\
\hline
\multicolumn{1}{c}{} &\multicolumn{1}{c}{A}  &\multicolumn{1}{c}{B} 
 &\multicolumn{1}{c}{C} &\multicolumn{1}{c}{D} \\
\hline	
\multicolumn{1}{c}{$V_{B} $} &\multicolumn{1}{c}{Energy}  &\multicolumn{1}{c}{Energy} 
 &\multicolumn{1}{c}{Energy} &\multicolumn{1}{c}{Energy} \\
\hline
-2.6047 &  -0.066674 & -0.066653 & -0.066653 & -0.06663 \\
-2.5 &  -0.310114 & -0.310115 & -0.310115 & -0.31004 \\
-2.3 &  -1.229845 & -1.229845 & -1.229845 &  -1.22970 \\
-2.1 &	-2.679069 & -2.678979 & -2.678979 & -2.67878 \\
\hline
\end{tabular}
\caption{Calculation of the neutron-proton virtual state 
 as function of interaction parameter
$\lambda $ for the $s$-wave Malfliet-Tjon interaction using the deformed 
integration contour $C_{2}^{+}$. The convergence is illustrated 
by increasing the number of integration points.
Column A used N1 = 30, N2 = 50 integration points, column B used N1 = 100, 
N2 = 100 integration points and  
column C used N1 = 150, N2 = 250 integration points. 
Comparison is made with the calculations of Ref.~\cite{elster}, given in column D.} 
\label{tab:tab6}
\end{table} 
We see a small difference in the calculated values for the virtual state in the
Malfliet-Tjon interaction. 
As we obtained convergence of the virtual state energies by increasing the number
of integration points, this suggests that our results are stable.
Fig.~\ref{fig:fig9} shows a plot of the calculated $t$-matrix elements
$t_{l}(k,k;E) $ for incoming energy $E = 100$ MeV for the $s$ -wave Malfliet-Tjon
interaction, using the contours $C_{1}^{+} $ and $C_{2}^{+}$. In this case we used a 
rotation angle $\theta = \pi/6 $ for both contours, and a translation $b = 
300$ MeV for contour $C_{2}^{+}$. The Malfliet-Tjon interaction has a 
singularity along contour $C_{1}^{+}$, given by $k_{max} = \mu /\tan(\theta ) = 
529.77$ MeV. In this case the interaction 
is not sufficiently neglible for $k,k' > k_{max}$,
and using contour $C_{1}^{+}$ will not give accurate calculation of the $t$-matrix,
this is clearly seen in the plot of the calculated results. 
This illustrates clearly the advantage of integrating along the contour $C_{2}^{+}$
instead of $C_{1}^{+}$. 

 A useful check 
of the numerics is the on-shell unitarity for the $S$-matrix. Table ~\ref{tab:tab7} reports
 calculations done by the principal-value prescription and the
contour deformation method $C_{2}^{+}$. We used $N=50 $ integration points for the principal value
integration, for the contour deformation method we used $N1 = 30$ and $N2 = 
100,200$.  We observe that we need a higher number of integration points
along line $L_{2}$ for obtaining convergence. 
\begin{table}[htbp]
  \begin{tabular}{rrrrr}
\hline
    \multicolumn{1}{c}{} & \multicolumn{1}{c}{PV} & \multicolumn{2}{c}{CDM} \\
\hline
    \multicolumn{1}{c}{$k$[MeV]} & \multicolumn{1}{c}{$N = 50$} & \multicolumn{1}{c}{$N2 = 100 $} & \multicolumn{1}{c}{$N2 = 200 $ }  \\
    \hline
    10. & 	1.00000000 & 	 1.00000811 & 	1.00000811 \\
    110. & 1.00000000 & 	 0.999999940 &	 1.00000000 \\
    210. & 1.00000000 & 	 0.999999940 &  1.00000000 \\
    310. & 1.00000000 &	0.999999881 &	1.00000000 \\
    410. & 1.00000000 &	 0.999999881 &	1.00000000 \\
    510. & 1.00000000 &	 0.999999881 & 	1.00000000 \\
    610. & 1.00000000 &	0.999999881 & 	1.00000000 \\
    710. & 1.00000000 & 	 0.999999821 &	1.00000000 \\
    810. & 1.00000000 & 	0.999999821 &	1.00000000 \\
    910. & 1.00000000 &	0.999999821 &	1.00000000 \\
    1010.& 1.00000000 &	0.999999762 &	1.00000000 \\
    1110.&  1.00000000& 	 0.999999702 &	 1.00000000 \\
    1210.&  1.00000000& 	 0.999999702 & 	1.00000000 \\
    1310.&  1.00000000& 	 0.999999642 & 	1.00000000 \\
    1410.&  1.00000000& 	 0.999999523 & 1.00000000 \\
    1510.&  1.00000000& 	 0.999999642 & 	1.00000000 \\
    \hline
  \end{tabular}
  \caption{Calculations of $S$-matrix norm, $\vert S(k) \vert $, for the $s$-wave
    Malfliet-Tjon potential with parameters  $ V_{A} = 7.291$ MeV, 
    $\mu_{A} = 613.69$ MeV, $V_{B} = -5$ MeV and  $\mu_{B} = 305.86$ MeV. Column 2 gives results for the
    principal value prescription (PV), while columns 3,4 give results for the contour
    deformation method (CDM).}
\label{tab:tab7}
\end{table}

\subsection{Results for the CD-Bonn nucleon-nucleon interaction}
Finally, we present a calculation of the energy spectrum of the 
charge-dependent Bonn interaction (CD-Bonn) and phase shifts for a set of selected
 partial waves. The CD-Bonn interaction is given in Ref.~\cite{machleidt}. 
We illustrate the power of the contour deformation method by reproducing the deuteron bound state and 
the virtual 
states in the ${}^{1}S_{0} $ isospin triplet channel. In addition, phase 
shifts are calculated within the same method, and the results agree perfectly with
other theoretical predictions. 

The tensor component of the CD-Bonn interaction
couples angular momentum, $l=j-1, l = j+1$,  in the spin triplet channel. 
It is straightforward to include this coupling in the formalism outlined
in the previous sections. Instead of $N$x$N $ matrices we get $2N$x$2N$ matrices
in the coupled case, for the uncoupled channels we still have $N$x$N$ 
matrices. We should of course label the equations with quantum numbers 
$L,S,J,T$. 

The realistic nucleon-nucleon interaction does not have a 
resonant structure in the low energy region $ E < 350$ MeV. 
It does however support virtual states in the ${}^{1}S_{0} $ isospin triplet
 channel and a bound state in the coupled isospin singlet channel (the deuteron) 
$ {}^{3}S_{1} - {}^{3}D_{1} $. We compare the calculated virtual state locations in 
the complex $k$-plane with the values obtained by the effective range 
approximation, see Ref.~\cite{newton,kukulin}. In the following we do not include Coulomb
effects when considering the isospin $t_{z} = -1 $ channel (proton-proton scattering).

The effective range approximation for the $s$-wave poles is given by
\begin{equation}
  \label{eq:eff_range}
  k = i \left[ {1\over r_{NN}} - \sqrt{ {2\over r_{NN}\vert a_{NN}\vert } + {1\over r_{NN}^{2} }}\right].
\end{equation}
The theoretical (see below) and experimental values \cite{machleidt} 
for the  ${}^{1}S_{0} $ scattering lengths $a_{NN}$ and 
effective range $r_{NN}$  are  given in Table~\ref{tab:tab8} 
\begin{table}[htbp]
\begin{tabular}{rrr}\hline
 \multicolumn{1}{c}{} & \multicolumn{1}{c}{CD-Bonn} & \multicolumn{1}{c}{Experiment}   \\
\hline
$a_{pp}$ & $-17.4602 $ & {} \\
$r_{pp}$ & 2.845 & {} \\
$a_{nn}$ & -18.9680 & -18.9 $\pm $0.4 \\
$r_{nn}$ & 2.819 & 2.75 $\pm $0.11 \\
$a_{np}$ & -23.7380 & -23.740 $\pm $0.020 \\
$r_{np}$ & 2.671 & 2.77 $\pm $0.05 \\ 
\hline
\end{tabular}
\caption{Scattering lengths (a) and effective ranges (r) for the  ${}^{1}S_{0} $ channel, 
  in units of fm. For the proton - proton channel Coulomb effects are not included.}
\label{tab:tab8}
\end{table}

To apply the contour deformation method by integrating along contour $C_{2}^{+}$ we 
first have to locate possible singularities in the CD-Bonn interaction.
The CD-Bonn interaction 
is given explicitly in momentum space. The derivation of the interaction is based
on field theory, starting from Lagrangians describing the coupling of the various 
mesons of interest to nucleons \cite{machleidt}.
 The one-boson-exchange interaction is proportional to 
\begin{equation}
  \label{eq:cdbonn}
  V({\bf k,q} ) \propto \sum _{\alpha = \pi^{0},\pi^{\pm}, \rho ,\omega,
    \sigma_{1},\sigma_{2}}\bar{V}_{\alpha}({\bf k,q } )F_{\alpha}^{2}({\bf k,q}; \Delta_{\alpha}).
\end{equation}
Both $\bar{V}_{\alpha}({\bf k,q } ) $ and $F_{\alpha}^{2}({\bf k,q}; \Delta_{\alpha}) $ 
contain terms of the form 
\begin{equation}
  {1\over ({\bf k -q } )^{2} + m_{\alpha }^{2} },
\end{equation}
which are of the Yukawa type. 
These terms determine the analyticity region of the interaction in the $k$-plane
(see discussion of the Malfliet-Tjon interaction above. 
We see that the poles of the interaction are determined  by the various meson masses 
$m_{\alpha}$ and cut-off masses $\Delta _{\alpha}$. 
Considering the solution of the eigenvalue problem by the 
contour deformation method using contour $C_{2}^{+}$, singularities in the interaction
appear for  $\theta \geq \pi/2 $, i.e., a rotation into the third quadrant of the
complex $k$-plane. The maximum translation into the complex $k$-plane is then determined 
by the smallest meson mass entering the potential, which is the $\pi $-meson, $m_{\pi^{0}} 
= 134.9764$ MeV. For a given rotation into the third quadrant, i.e., $\theta \geq \pi/2 $,
 we get a restriction on the translation parameter; $ b < 134.9764/2\sin(\theta )$. Here $b$ is given in MeV, as
we are still using natural units.

Table~\ref{tab:tab9} gives results for the virtual states in the 
 ${}^{1}S_{0} $ channel by the contour deformation method. A comparison with
the effective range calculation of the virtual state poles is also shown. 
The contour was
rotated by $\theta = 2\pi /3 $ into the complex $k$-plane and translated 
$c = -30\sin(5\pi/7)$ MeV or $c\approx -23.45$ MeV in the lower-half $k$-plane. 
This is sufficient to reproduce the virtual states in the ${}^{1}S_{0} $ channel, as they are 
known to lie very close to the scattering threshold, $k\approx -\hbar c 0.05i$ MeV. 
By this contour choice the full energy spectrum is obtainable since it is known 
\emph{a posteriori} that the ${}^{1}S_{0} $ channel supports only virtual states near 
the scattering threshold. 
The convergence of the
numerical calculated values is demonstrated by increasing the number of integration points $N$. 
\begin{table}[htbp]
\begin{tabular}{rrrrr}\hline
  \multicolumn{1}{c}{} & \multicolumn{3}{c}{CDM}  &\multicolumn{1}{c}{EFR}   \\
  \hline
  \multicolumn{1}{c}{} & \multicolumn{1}{c}{a} & \multicolumn{1}{c}{b} & \multicolumn{1}{c}{c}
  & \multicolumn{1}{c}{} \\  
  \hline
  \multicolumn{1}{c}{$T_{z}$} & \multicolumn{1}{c}{Energy} & 
  \multicolumn{1}{c}{Energy} & \multicolumn{1}{c}{Energy} &
  \multicolumn{1}{c}{Energy}  \\
  \hline
  -1 (pp) &  -0.11766     &   -0.11761   &   -0.11761   &  -0.11763 \\
  1 (nn)  &  -0.10070     &   -0.10069   &   -0.10069   &  -0.10070 \\
  0 (np)  &  -0.06632     &   -0.06632   &   -0.06632   &  -0.06632 \\
  \hline
\end{tabular}
\caption{Calculation of virtual state energies in the ${}^{1}S_{0}$ isospin triplet channel 
 by the effective range approximation (EFR) and the contour deformation method (CDM), 
in units of MeV.
Convergence is obtained in each isospin channel. Column (a) used $N1 = 20, N2 = 30$ integration points,
column (b) used  $N1=20, N2 = 50$ integration points while column (c) used 
$N1 = 30, N2 = 80$ integration points }
\label{tab:tab9}
\end{table}
Table~\ref{tab:tab10} illustrates the convergence of the deuteron binding energy as
a function of integration points. The contour was rotated by $\theta = \pi /6 $ into the complex $k$-plane and translated 
$c = -\hbar c\sin(\pi/6)$ MeV or $ c\approx -100$ MeV into the lower-half $k$-plane. 
\begin{table}[htbp]
\begin{tabular}{rrr}\hline
  \multicolumn{1}{c}{N1} & \multicolumn{1}{c}{N2} & \multicolumn{1}{c}{Energy} \\ 
  \hline
  20 & 30 &  -2.224581 \\
  20 & 50 &  -2.224573 \\  
  30 & 80 &  -2.224574 \\
  50 & 100 & -2.224575 \\ 
  50 & 150 & -2.224575 \\
  \hline
\end{tabular}
\caption{Calculation of deuteron binding energy in the coupled channel 
  ${}^{3}S_{1}-{}^{3}D_{1}$  by the contour deformation method (CDM), in units
of MeV. See Ref.~\cite{machleidt} for further details.}
\label{tab:tab10}
\end{table}

\begin{figure}[hbtp]
\begin{center}
\resizebox{7cm}{7cm}{\epsfig{file=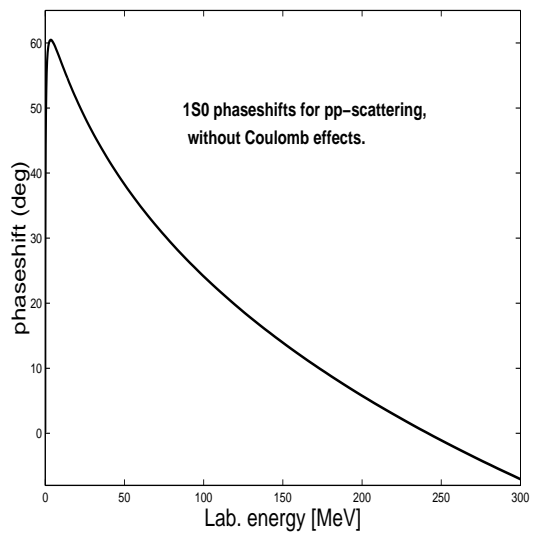}}
\end{center} 
\caption{Phaseshift for proton-proton scattering in the ${}^{1}S_{0}$ channel. Coulomb 
effects are not included in the calculations. }
\label{fig:pp}
\end{figure}

\begin{figure}[hbtp]
\begin{center}
\resizebox{7cm}{7cm}{\epsfig{file=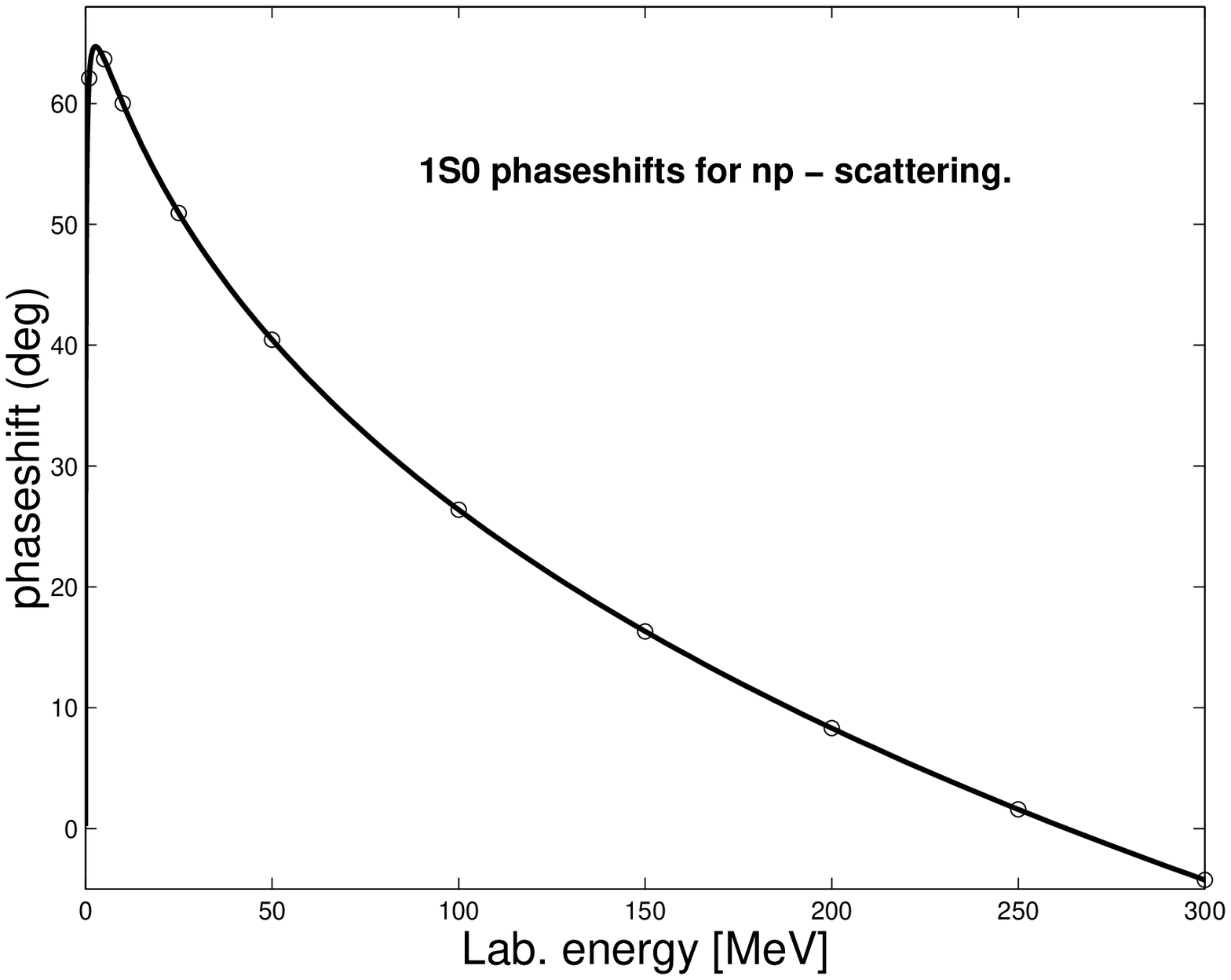}}
\end{center}
\caption{Phaseshift for neutron-proton scattering in the ${}^{1}S_{0}$ channel. 
Calculation by the contour deformation method is given by the solid line. Circles are
calculations given in Ref.~\cite{machleidt}. }
\label{fig:np}
\end{figure}

\begin{figure}[hbtp]
\begin{center}
\resizebox{7cm}{7cm}{\epsfig{file=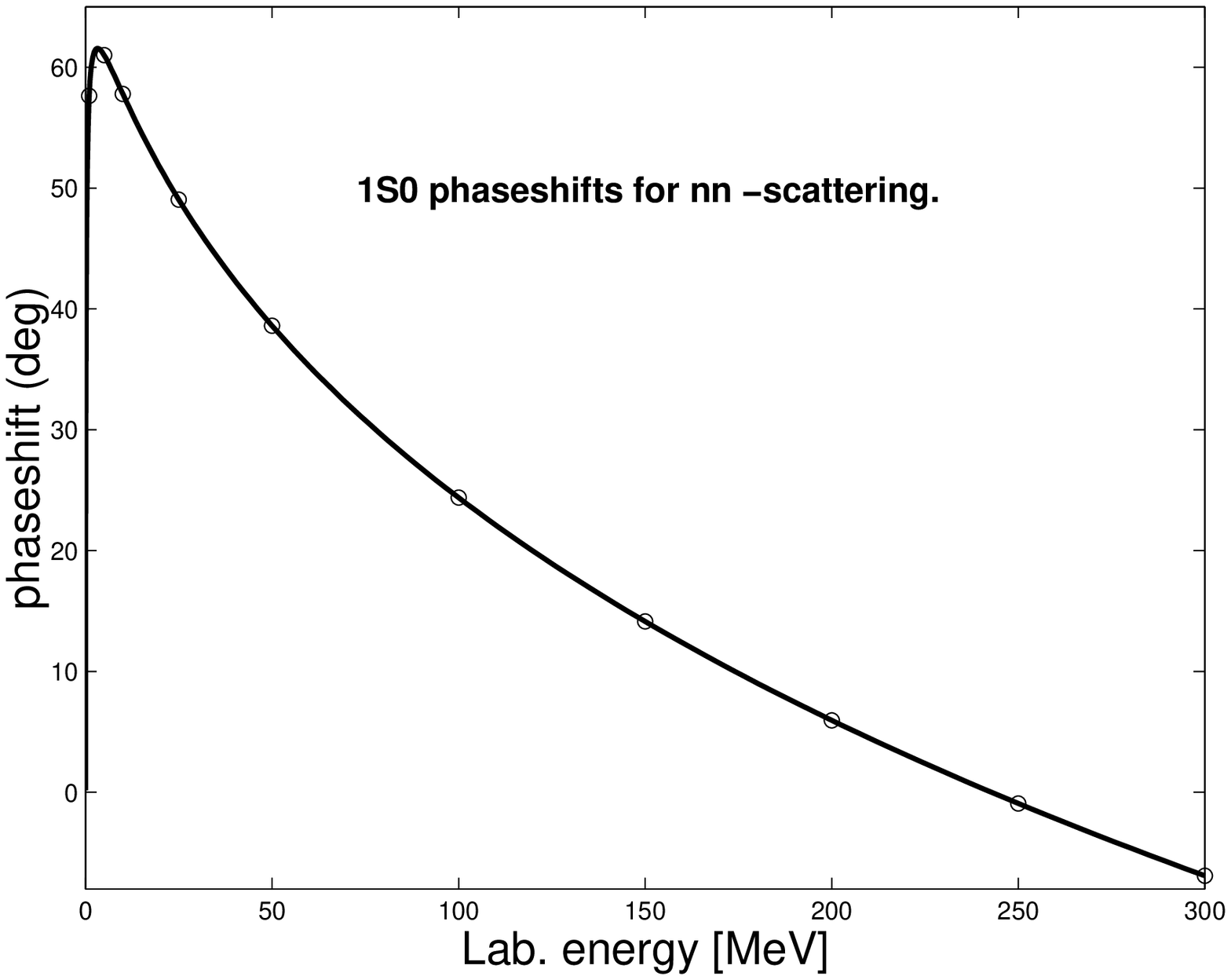}}
\end{center}
\caption{Phaseshift for neutron-proton scattering in the ${}^{1}S_{0}$ channel. 
Calculation by the contour deformation method is given by the solid line. Circles are
calculations given in Ref.~\cite{machleidt}. }
\label{fig:nn}
\end{figure}

Figs.~\ref{fig:pp},\ref{fig:np} and \ref{fig:nn} present calculations of 
nucleon-nucleon phaseshifts for 
the uncoupled ${}^{1}S_{0}$ isospin triplet channel, by the contour 
deformation method. The contour was rotated by $2\pi /3 $ into the third quadrant of
the complex $k$-plane, and translated $c = 50\sin(2\pi/3)$ MeV. We used $N1 = 30 $ and
$ N2 = 70  $ integration points along line $L_{1}$ and line $L_{2}$, respectively. 
Our calculation is given by 
the solid line while the circles are calculations given in Ref.~\cite{machleidt}. 
For proton-proton scattering Coulomb effects are not included. 
        
\section{Conclusions and perspectives}

A generalized contour deformation method in momentum space has been presented.
The deformation of the integration contour is generalized to rotation followed
by translation in the complex $k$-plane. This generalization makes it 
possible to handle both \emph{dilation} and \emph{non-dilation} invariant
potentials. We have shown this to be a powerful procedure for studying
resonances and virtual states in two-body systems. The method has also 
been successfully applied to the calculation of the full off-shell $t$-matrix.
The aim of this work has been to establish the formalism for the free scattering case,
basing the analysis on schematic and realistic nucleon-nucleon interactions.
Work on extension to few-body Borromean halo systems \cite{zhukov} with more than two constituents, is 
in progress.

The exposed formalism allows for stable numerical calculations 
of bound states, resonances and
virtual states, in addition to yielding a fully complex on-shell and off-shell scattering matrix,
starting with a realistic nucleon-nucleon interaction.

This allows for several interesting applications and extensions.  
As is well-known, one of the major challenges in the microscopic description of weakly bound nuclei is
a proper treatment of both the many-body correlations and the continuum of positive energies
and decay channels. Such nuclei pose a tough challenge on traditional nuclear structure methods,
based essentially on the derivation of effective interactions and the nuclear shell-model, see
for example Ref.~\cite{mhj95} for a review. In the traditional approaches only bound states typically 
enter the determination of an effective interaction, be it either based upon various many-body schemes 
or more phenomenologically  inspired approaches. Coupled with large-scale shell model studies, several 
properties of stable nuclei are well reproduced. However, weakly bound nuclei may have a strong coupling to 
unbound states, either resonances or virtual states, as described in for example  
Refs.~\cite{witek1, witek2, roberto}.  This implies in turn that an effective interaction should reflect
such couplings with the continuum, i.e., a consistent many-body scheme should include bound states, 
resonances and virtual states as well.
 
The present approach may allow for such a scheme. The simplest extension is to consider 
in-medium scattering of two nucleons in e.g., infinite nuclear matter or neutron star matter, as done 
by Dickhoff and {\em et al.}  \cite{dickhoff1998} or Schulze {\em et al.} \cite{hansjosef}.
This is easily accomplished by inserting an exclusion operator $Q$ in Eq.~(\ref{eq:spectral1}) 
for the Green's operator. This exclusion operator can be constructed so as to prevent scattering 
into occupied states. Typically, one can then generate selected hole-hole, particle-hole 
and particle-particle intermediate
state correlations that reflect a given nuclear medium. Combined with a self-consistent determination of the 
single-particle energies it is then possible to derive various classes  of diagrams at the two-body level.
Such work is in progress, based on the present method and the exact definition of the 
exclusion operator $Q$ by Suzuki {\em et al.} \cite{kenyj_rioyj}.


\begin{thebibliography}{200}
\bibitem{newton} R.~G.~Newton \emph{Scattering Theory of Waves and Particles} 
(Springer-Verlag, New York, 1966,1982). 
\bibitem{kukulin} V.~I.~Kukulin, V.~M.~Krasnopol'sky, and 
J.~Hor$\mathrm{\acute{a}\check{c}}$ek,
\emph{Theory of Resonances} (Kluwer Academic publishers 1989).
\bibitem{afnan1} I.~R.~Afnan, Aust.~J.~Phys.~{\bf 44 }, 201 (1991).
\bibitem{abc} J.~Aguilar and J.~M.~Combes, Commun.~Math.~Phys.~{\bf 22}, 269 (1971).
\bibitem{abc1} E.~Balslev and J.~M.~Combes, Commun.~Math.~Phys.~{\bf 22}, 280 (1971). 
\bibitem{moise} N.~Moiseyev, Phys.~Rep.~{\bf 302}, 211 (1998). 
\bibitem{csoto} A.~Csoto, Phys.~Rev.~C {\bf 49}, 2244 (1994).
\bibitem{garrido} E.~Garrido, D.~V.~Fedorov, and A.~S.~Jensen, 
	Nucl.~Phys.~A {\bf 708}, 277 (2002). 
\bibitem{imante} I. Raskinyte, \emph{Resonances in few-body systems}, 
PhD Thesis University of Bergen 2002.
\bibitem{brayshaw} D.~Brayshaw, Phys.~Rev.~{\bf 176 }, 1855 (1968).
\bibitem{nuttal} J.~Nuttall and H.~L.~Cohen, Phys.~Rev.~{\bf 188}, 1542 (1969).
\bibitem{stelbovics} A.~T.~Stelbovics, Nucl.~Phys.~A {\bf 288}, 461 (1978).
\bibitem{berggren} T.~Berggren, Nucl.~Phys.~A {\bf 109}, 265 (1968)
\bibitem{lind} P.~Lind, Phys.~Rev.~C {\bf 47}, 1903 (1993)
\bibitem{nuttal1} J.~Nuttall, J.~Math.~Phys.~{\bf 8}, 873 (1967).
\bibitem{tikto} G.~Tiktopoulos, Phys.~Rev.~{\bf 136}, 275 (1964).
\bibitem{aoyama} S.~Aoyama, Phys.~Rev.~Lett.~{\bf 89}, 052501 (2002).
\bibitem{afnan} B.~C.~Pearce and I.~R.~Afnan, Phys.~Rev.~C {\bf 30}, 2022 (1984).
\bibitem{yamaguchi} K.~Yamaguchi, Phys.~Rev.~{\bf 95}, 1628 (1954).
\bibitem{malfliet} R.~A.~Malfliet and J.~A.~Tjon, Nucl.~Phys.~A {\bf 127}, 161 (1969).
\bibitem{machleidt} R.~Machleidt, Phys.~Rev.~C {\bf 63}, 024001 (2001).
\bibitem{betan} R.~Id Betan, R.~J.~Liotta, N.~Sandulescu, and T.~Vertse, 
		Phys.~Rev.~C {\bf 67}, 014322 (2003).
\bibitem{brown} L.~Brown, D.~Fivel, B.~W.~Lee, and R.~Sawyer, Ann.~of Phys.~{\bf 23}, 167 (1963).
\bibitem{elster} Ch.~Elster, J.~H.~Thomas, and W.~Gl\"ockle, Few-Body Systems {\bf 24}, 55 (1998).
\bibitem{zhukov} M.~V.~Zhukov, B.~V.~Danilin, D.~V.~Fedorov, J.~M.~Bang, 
I.~J.~Thompson, and J.~S.~Vaagen, Phys.~Rep.~{\bf 231}, 151 (1993). 
\bibitem{mhj95} M.~Hjorth-Jensen, T.~T.~S.~Kuo, and E.~Osnes, Phys.~Rep.~{\bf 261}, 
125 (1995).
\bibitem{witek1} N.~Michel, W.~Nazarewicz, M.~P{\l}oszajczak, and K.~Bennaceur, 
                 Phys.~Rev.~Lett.~{\bf 89}, 042502 (2002).
\bibitem{witek2} N.~Michel, W.~Nazarewicz, M.~P{\l}oszajczak, and J.~Oko{\l}owicz, nucl-th/0302060.
\bibitem{roberto}R.~Id Betan, R.~J.~Liotta, N.~Sandulescu, and T.~Vertse, 
                 Phys.~Rev.~Lett.~{\bf 89}, 042501 (2002).
\bibitem{dickhoff1998} W.~H.~Dickhoff, Phys.~Rev.~C {\bf 58}, 2807 (1998); W.~H.~Dickhoff, C.~C.~Gearhart, 
                   E.~P.~Roth, A.~Polls, and A.~Ramos, Phys.~Rev.~C {\bf 60}, 064319 (1999). 
\bibitem{hansjosef} H.-J.~Schulze, A.~Schnell, G.~R\"opke, and U.~Lombardo, Phys.~Rev.~C {\bf 55}, 3006 (1997). 
\bibitem{kenyj_rioyj} K.~Suzuki, R.~Okamoto, M.~Kohno, and S.~Nagata, Nucl.~Phys.~A {\bf 665}, 92 (2000).
\end{thebibliography}
\end{document}